\documentclass[aps,prl,twocolumn,superscriptaddress,nofootinbib,floatfix]{revtex4-2}

\usepackage[dvipsnames]{xcolor}
\usepackage{framed}
\definecolor{shadecolor}{rgb}{0.9,0.9,0.9}

\usepackage{mathtools}
\usepackage{amsmath}
\usepackage[shortlabels]{enumitem}

\usepackage{graphicx,epic,eepic,epsfig,amsmath,latexsym,amssymb,verbatim,color}
 
\usepackage{amsfonts}       
\usepackage{nicefrac}       

\usepackage{amsmath}
\usepackage{bbm}

\usepackage{tikz}
\usetikzlibrary{chains}
\usetikzlibrary{fit}
\usetikzlibrary{arrows}
\usetikzlibrary{decorations}

\usepackage{epsfig}
\usetikzlibrary{shapes.symbols,patterns} 
\usepackage{pgfplots}

\usepackage[strict]{changepage}
\usepackage{hyperref}
\hypersetup{colorlinks=true,citecolor=blue,linkcolor=blue,filecolor=blue,urlcolor=blue,breaklinks=true}

\usepackage[marginal]{footmisc}
\usepackage{url}
\usepackage{theorem}
 
\def\squareforqed{\hbox{\rlap{$\sqcap$}$\sqcup$}}
\def\qed{\ifmmode\squareforqed\else{\unskip\nobreak\hfil
\penalty50\hskip1em\null\nobreak\hfil\squareforqed
\parfillskip=0pt\finalhyphendemerits=0\endgraf}\fi}
\def\endenv{\ifmmode\;\else{\unskip\nobreak\hfil
\penalty50\hskip1em\null\nobreak\hfil\;
\parfillskip=0pt\finalhyphendemerits=0\endgraf}\fi}

\newcounter{remark}

\newcounter{example}

\mathchardef\ordinarycolon\mathcode`\:
\mathcode`\:=\string"8000
\def\vcentcolon{\mathrel{\mathop\ordinarycolon}}
\begingroup \catcode`\:=\active
  \lowercase{\endgroup
  \let :\vcentcolon
  }

\usepackage{cleveref}
\usepackage{graphicx}
\usepackage{xcolor}

\RequirePackage[framemethod=default]{mdframed}
\newmdenv[skipabove=7pt,
skipbelow=7pt,
backgroundcolor=darkblue!15,
innerleftmargin=5pt,
innerrightmargin=5pt,
innertopmargin=5pt,
leftmargin=0cm,
rightmargin=0cm,
innerbottommargin=5pt,
linewidth=1pt]{tBox}

\newmdenv[skipabove=7pt,
skipbelow=7pt,
backgroundcolor=red!15,
innerleftmargin=5pt,
innerrightmargin=5pt,
innertopmargin=5pt,
leftmargin=0cm,
rightmargin=0cm,
innerbottommargin=5pt,
linewidth=1pt]{rBox}

\newmdenv[skipabove=7pt,
skipbelow=7pt,
backgroundcolor=blue2!25,
innerleftmargin=5pt,
innerrightmargin=5pt,
innertopmargin=5pt,
leftmargin=0cm,
rightmargin=0cm,
innerbottommargin=5pt,
linewidth=1pt]{dBox}
\newmdenv[skipabove=7pt,
skipbelow=7pt,
backgroundcolor=darkkblue!15,
innerleftmargin=5pt,
innerrightmargin=5pt,
innertopmargin=5pt,
leftmargin=0cm,
rightmargin=0cm,
innerbottommargin=5pt,
linewidth=1pt]{sBox}
\definecolor{darkblue}{RGB}{0,76,156}
\definecolor{darkkblue}{RGB}{0,0,153}
\definecolor{blue2}{RGB}{102,178,255}
\definecolor{darkred}{RGB}{195,0,0}

\newcommand{\nc}{\newcommand}
\nc{\rnc}{\renewcommand}
\nc{\lbar}[1]{\overline{#1}}
\nc{\bra}[1]{\langle#1|}
\nc{\cptp}{\operatorname{CPTP}}
\nc{\hptp}{\operatorname{HPTP}}
\nc{\ket}[1]{|#1\rangle}
\nc{\spn}{\operatorname{span}}

\nc{\rep}[1]{\textbf{R}(#1)}

\nc{\ketbra}[2]{|#1\rangle\!\langle#2|}
\nc{\dketbra}[2]{|#1\rangle\!\rangle\!\langle\!\langle#2|}
\nc{\dket}[1]{|#1\rangle\!\rangle}
\nc{\dbra}[1]{\langle\!\langle#1|}
\nc{\dbraket}[2]{\langle\!\langle#1|#2\rangle\!\rangle}

\nc{\braket}[2]{\langle#1|#2\rangle}

\nc{\proj}[1]{| #1\rangle\!\langle #1 |}
\nc{\avg}[1]{\langle#1\rangle}
\nc{\smfrac}[2]{\mbox{$\frac{#1}{#2}$}}
\nc{\tr}{\operatorname{Tr}}
\nc{\ox}{\otimes}
\nc{\dg}{\dagger}
\nc{\dn}{\downarrow}
\nc{\cA}{{\cal A}}
\nc{\cB}{{\cal B}}
\nc{\cC}{{\cal C}}
\nc{\cD}{{\cal D}}
\nc{\cE}{{\cal E}}
\nc{\cF}{{\cal F}}
\nc{\cG}{{\cal G}}
\nc{\cH}{{\cal H}}
\nc{\cI}{{\cal I}}
\nc{\cJ}{{\cal J}}
\nc{\cK}{{\cal K}}
\nc{\cL}{{\cal L}}
\nc{\cM}{{\cal M}}
\nc{\cN}{{\cal N}}
\nc{\cO}{{\cal O}}
\nc{\cP}{{\cal P}}
\nc{\cQ}{{\cal Q}}
\nc{\cR}{{\cal R}}
\nc{\cS}{{\cal S}}
\nc{\cT}{{\cal T}}
\nc{\cU}{{\cal U}}
\nc{\cV}{{\cal V}}
\nc{\cX}{{\cal X}}
\nc{\cY}{{\cal Y}}
\nc{\cZ}{{\cal Z}}
\nc{\cW}{{\cal W}}
\nc{\csupp}{{\operatorname{csupp}}}
\nc{\qsupp}{{\operatorname{qsupp}}}
\nc{\var}{{\operatorname{var}}}
\nc{\rar}{\rightarrow}
\nc{\lrar}{\longrightarrow}
\nc{\polylog}{{\operatorname{polylog}}}
\nc{\wt}{{\operatorname{wt}}}
\nc{\av}[1]{{\left\langle {#1} \right\rangle}}
\nc{\supp}{{\operatorname{supp}}}

\nc{\argmin}{{\operatorname{argmin}}}

\def\x{\xi}

\nc{\RR}{{{\mathbb R}}}
\nc{\CC}{{{\mathbb C}}}
\nc{\FF}{{{\mathbb F}}}
\nc{\NN}{{{\mathbb N}}}
\nc{\ZZ}{{{\mathbb Z}}}
\nc{\PP}{{{\mathbb P}}}
\nc{\QQ}{{{\mathbb Q}}}
\nc{\UU}{{{\mathbb U}}}
\nc{\EE}{{{\mathbb E}}}
\nc{\id}{{\operatorname{id}}}

\nc{\CHSH}{{\operatorname{CHSH}}}

\nc{\be}{\begin{equation}}
\nc{\ee}{{\end{equation}}}
\nc{\bea}{\begin{eqnarray}}
\nc{\eea}{\end{eqnarray}}
\nc{\<}{\langle}
\rnc{\>}{\rangle}
\nc{\rU}{\mbox{U}}

\nc{\ob}[1]{#1}

\nc{\SEP}{{\text{\rm SEP}}}
\nc{\NS}{{\text{\rm NS}}}
\nc{\LOCC}{{\text{\rm LOCC}}}
\nc{\PPT}{{\text{\rm PPT}}}
\nc{\EXT}{{\text{\rm EXT}}}
\nc{\Sym}{{\operatorname{Sym}}}

\nc{\ERLO}{{E_{\text{r,LO}}}}
\nc{\ERLOCC}{{E_{\text{r,LOCC}}}}
\nc{\ERPPT}{{E_{\text{r,PPT}}}}
\nc{\ERLOCCinfty}{{E^{\infty}_{\text{r,LOCC}}}}
\nc{\Aram}{{\operatorname{\sf A}}}

\usepackage{tikz}
\hypersetup{colorlinks=true,citecolor=blue,linkcolor=blue,filecolor=blue,urlcolor=blue,breaklinks=true}

\makeatletter
\def\grd@save@target#1{%
  \def\grd@target{#1}}
\def\grd@save@start#1{%
  \def\grd@start{#1}}
\tikzset{
  grid with coordinates/.style={
    to path={%
      \pgfextra{%
        \edef\grd@@target{(\tikztotarget)}%
        \tikz@scan@one@point\grd@save@target\grd@@target\relax
        \edef\grd@@start{(\tikztostart)}%
        \tikz@scan@one@point\grd@save@start\grd@@start\relax
        \draw[minor help lines,magenta] (\tikztostart) grid (\tikztotarget);
        \draw[major help lines] (\tikztostart) grid (\tikztotarget);
        \grd@start
        \pgfmathsetmacro{\grd@xa}{\the\pgf@x/1cm}
        \pgfmathsetmacro{\grd@ya}{\the\pgf@y/1cm}
        \grd@target
        \pgfmathsetmacro{\grd@xb}{\the\pgf@x/1cm}
        \pgfmathsetmacro{\grd@yb}{\the\pgf@y/1cm}
        \pgfmathsetmacro{\grd@xc}{\grd@xa + \pgfkeysvalueof{/tikz/grid with coordinates/major step}}
        \pgfmathsetmacro{\grd@yc}{\grd@ya + \pgfkeysvalueof{/tikz/grid with coordinates/major step}}
        \foreach \x in {\grd@xa,\grd@xc,...,\grd@xb}
        \node[anchor=north] at (\x,\grd@ya) {\pgfmathprintnumber{\x}};
        \foreach \y in {\grd@ya,\grd@yc,...,\grd@yb}
        \node[anchor=east] at (\grd@xa,\y) {\pgfmathprintnumber{\y}};
      }
    }
  },
  minor help lines/.style={
    help lines,
    step=\pgfkeysvalueof{/tikz/grid with coordinates/minor step}
  },
  major help lines/.style={
    help lines,
    line width=\pgfkeysvalueof{/tikz/grid with coordinates/major line width},
    step=\pgfkeysvalueof{/tikz/grid with coordinates/major step}
  },
  grid with coordinates/.cd,
  minor step/.initial=.2,
  major step/.initial=1,
  major line width/.initial=2pt,
}
\makeatother

\usepackage{thmtools}
\usepackage{thm-restate}
\usepackage{etoolbox}
\makeatletter
\def\problem@s{}
\newcounter{problems@cnt}

\newcommand{\allproblems}{\problem@s}
\makeatother
\usepackage{amsmath,amsfonts,amssymb,array,graphicx,mathtools,multirow,bm,tcolorbox,relsize,booktabs}
\usepackage[shortlabels]{enumitem}
\usepackage{graphicx,epic,times,eepic,epsfig,latexsym,verbatim,color}
\usepackage{nicefrac}       
\usepackage[linesnumbered,ruled,procnumbered]{algorithm2e}
\usepackage[table]{xcolor}
\usepackage{mathrsfs}
\usepackage{framed}

\definecolor{shadecolor}{rgb}{0.9,0.90,0.9}
\usepackage{bbm}
\usepackage{epsfig}
\usepackage{pgfplots}
\usepackage[strict]{changepage}
\usepackage{hyperref}
\hypersetup{colorlinks=true,citecolor=blue,linkcolor=blue,filecolor=blue,urlcolor=blue,breaklinks=true}
\usepackage{url}
\usepackage{theorem}
\pgfplotsset{compat=1.18} 

\usepackage[normalem]{ulem}
\usepackage{thmtools}
\usepackage{thm-restate}
\usepackage{etoolbox}
\usepackage[utf8]{inputenc}
\usepackage[T1]{fontenc}
\usepackage{quantikz} %

\allowdisplaybreaks
\begin{document}
\title{Phase-stable hologram updates for large-scale neutral-atom array reconfiguration}
\author{Erdong Huang}
\affiliation{Thrust of Artificial Intelligence, Information Hub,\\
The Hong Kong University of Science and Technology (Guangzhou), Guangzhou 511453, China}
\affiliation{QudeLeap Research, Shanghai 200030, China}
\author{Jiayi Huang}
\affiliation{Thrust of Artificial Intelligence, Information Hub,\\
The Hong Kong University of Science and Technology (Guangzhou), Guangzhou 511453, China}
\author{Hongshun Yao}
\affiliation{Thrust of Artificial Intelligence, Information Hub,\\
The Hong Kong University of Science and Technology (Guangzhou), Guangzhou 511453, China}
\affiliation{QudeLeap Research, Shanghai 200030, China}
\author{Xin Wang}
\email{felixxinwang@hkust-gz.edu.cn}
\affiliation{Thrust of Artificial Intelligence, Information Hub,\\
The Hong Kong University of Science and Technology (Guangzhou), Guangzhou 511453, China}

\author{Jin-Guo Liu}
\email{jinguoliu@hkust-gz.edu.cn}
\affiliation{Thrust of Advanced Materials, Function Hub,\\
The Hong Kong University of Science and Technology (Guangzhou), Guangzhou 511453, China}

\begin{abstract}
Dynamic holographic optical tweezers provide a programmable route to array assembly and reconfiguration essential for scalable neutral-atom quantum computation. However, phase mismatch between successive holograms can cause destructive interference during finite spatial light modulator (SLM) refresh. In this work, we analyze finite SLM refresh to establish a phase-stability criterion for prescribed trap amplitudes and develop a weighted-projective Gerchberg--Saxton (WPGS) method that efficiently approximates the corresponding phase-only complex-field optimization. Enforcing this phase constraint also reduces the number of iterations required for each update, achieving convergence within five iterations and enabling hologram generation within a few milliseconds. Numerical simulations of 2D and 3D reconfiguration involving more than \(10^3\) traps, together with nonuniform-intensity interlayer transport, show that WPGS preserves endpoint intensity quality while suppressing inter-frame phase mismatch, transient intensity degradation, trap splitting, and motional heating. Similar suppression of phase mismatch and transient-intensity degradation is maintained under Gaussian SLM-plane illumination. These results establish the phase-stability criterion as a practical design principle for dynamic holographic control and scalable neutral-atom array reconfiguration.
\end{abstract}

\maketitle


\textit{Introduction.}---
Quantum computation with neutral atoms has advanced rapidly, establishing
optical-tweezer arrays as a promising platform for scalable quantum
information processing and fault-tolerant quantum computing.
Realizing this potential requires reliable preparation, control,
manipulation, and measurement of large qubit arrays, together with
flexible connectivity and programmable interactions
~\cite{altman2021quantum,Tian2023Parallel,
Pichard2024Rearrangement,Schymik2020Enhanced,manetsch2025tweezer,
norcia2023midcircuit}.
These requirements become increasingly stringent in the fault-tolerant
regime, where large physical-qubit counts and substantial encoding
overheads place demanding constraints on array size, connectivity, and
controllability
~\cite{Pecorari2025LowDepth,bluvstein2024logical}.

Optical-tweezer arrays provide a powerful route toward meeting these
requirements
~\cite{kaufman2021QuantumScienceOptical,browaeys2020Manybody}.
They combine long coherence times and high-fidelity state preparation,
control, and readout with reconfigurable geometries and tunable
interactions
~\cite{Levine2018HighFidelity,Bernien2017Probingmany-body}.
These capabilities have enabled rapid progress in quantum simulation
and quantum information processing
~\cite{Bernien2017Probingmany-body,Ebadi2021Quantum,
semeghini2021probing,Levine2019Parallel,Henriet2020Quantumcomputing},
including high-fidelity parallel entangling gates, continuous operation
of thousand-qubit-scale arrays, programmable studies of nonequilibrium
and topological many-body physics, and the development of
fault-tolerant neutral-atom architectures
~\cite{Evered2023HighFidelity,Chiu2025Continuous,
Manovitz2025QuantumCoarsening,Evered2025Kitaev,
Bluvstein2026FaultTolerant,muniz2025repeated}.

A central challenge in scaling these systems is the deterministic
preparation and reconfiguration of large, defect-free atom arrays.
Because single-site loading is stochastic and leaves vacancies
~\cite{Barredo2016atombyatom,Brown2019GrayMolasses,
Shaw2023DarkState},
target arrays are typically assembled by atom rearrangement and
transport
~\cite{Barredo2016atombyatom,Lin2025AIEnabledRapidAssembly,
Pause2024Supercharged,Gyger2024Continuous}.
Phase-only spatial light modulators (SLMs) provide a versatile route to
holographic multitweezer generation, supporting large arrays, flexible
target geometries, dynamic reconfiguration, and multiple axial planes
with a single optical element
~\cite{nogrette2014single,tamura2016highly,
Chew2024Ultraprecise,Schlosser2023Scalable,
kim2016Situ,Lee2016Three-dimensional,machu2025fullfieldofview}.

Dynamic hologram updates, however, pose a qualitatively different
problem from static hologram generation. A liquid-crystal SLM switches
between phase patterns over a finite response time, during which the
optical field evolves coherently between the initial and final
holograms. Thus, two holograms with comparable endpoint intensity
fidelity can exhibit very different transient behavior: mismatched
trap-resolved optical phases can produce destructive interference
during the refresh and strongly reduce the instantaneous trapping
intensity. Conventional weighted Gerchberg--Saxton (WGS)-type methods
primarily optimize endpoint intensities and therefore do not, in
general, control the inter-frame phase relation governing this
transient interference
~\cite{Gerchberg1972APA,DiLeonardo:07,Pasienski2008,
nogrette2014single}.
Recent approaches have incorporated phase information through
AI-enabled joint position--phase control
~\cite{Lin2025AIEnabledRapidAssembly}
or prescribed interpolation of trap positions and phases
~\cite{Knottnerus2025Parallel}.
A general criterion and optimization framework for generating
successive phase-only holograms that simultaneously preserve prescribed
trap amplitudes and suppress refresh-induced transient degradation,
however, remains lacking.

\begin{figure}
    \centering
    \includegraphics[width=1\linewidth]{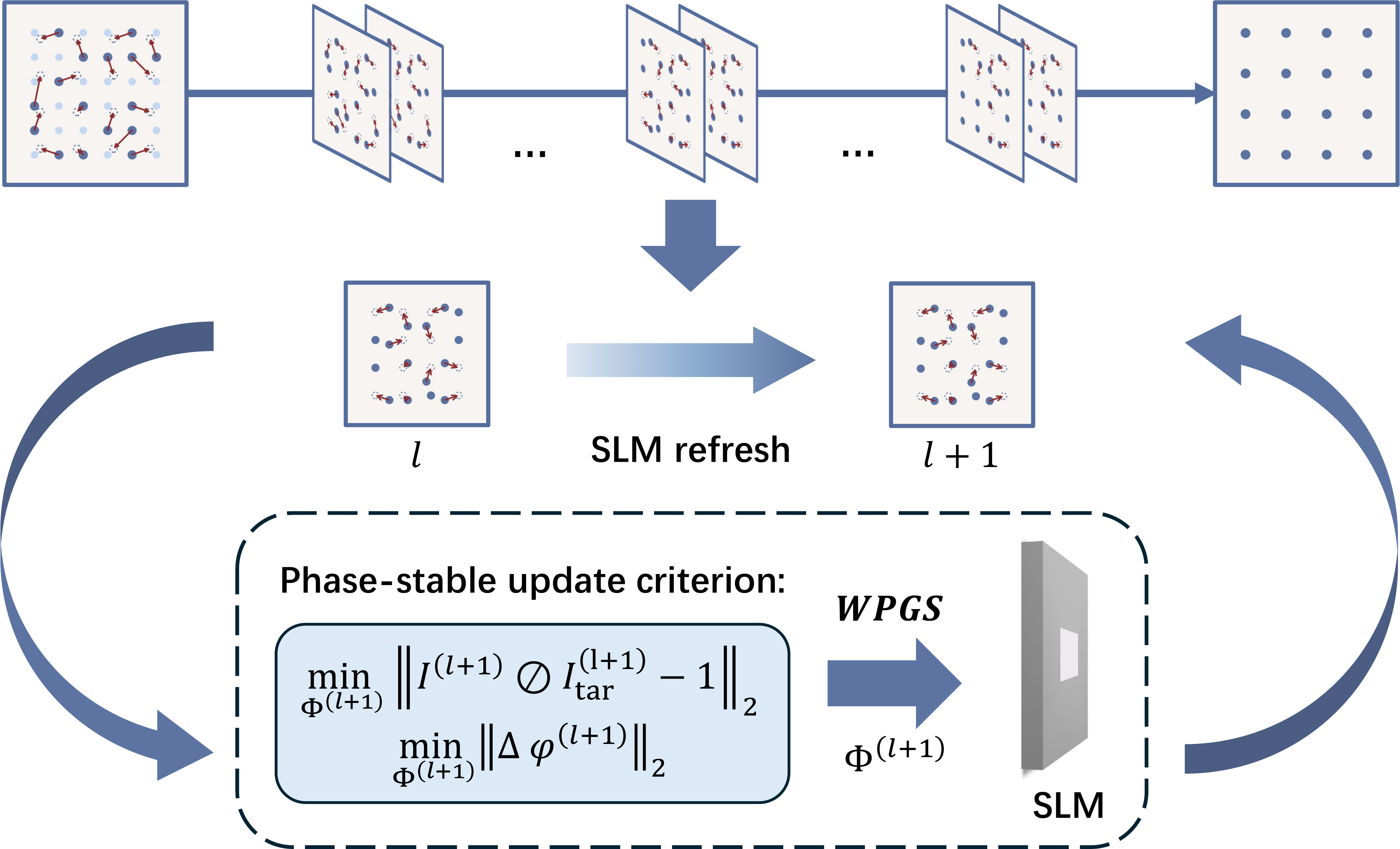}
\caption{\textbf{WPGS for refresh-stable dynamic hologram generation.}
A prescribed sequence of holograms generates optical-tweezer configurations from the initial to the target configuration. For each update \(l\to l+1\), the update criterion combines matching of the prescribed target intensities with inter-frame phase continuity. WPGS implements this criterion to obtain the phase-only SLM pattern \(\Phi^{(l+1)}\), suppressing transient intensity dips during finite SLM refresh.}
    \label{fig:schematic}
\end{figure}

In this work, we formulate a \emph{phase-stability criterion} for finite-time
SLM refresh under prescribed trap amplitudes and introduce
weighted-projective Gerchberg--Saxton (WPGS) optimization for
phase-stable hologram updates. For each target update, WPGS combines
the prescribed trap amplitudes with the realized trap phases of the
preceding frame and constructs a phase-only hologram by iteratively
matching the resulting complex target field, as illustrated in
Fig.~\ref{fig:schematic}. We demonstrate the method for large-scale
two-dimensional rearrangement, simultaneous three-dimensional
reconfiguration, and partial interlayer transport with independently
prescribed nonuniform trap intensities. In all three settings, WPGS
retains high endpoint intensity fidelity while strongly reducing
inter-frame phase mismatch and the associated transient loss of
trapping intensity. These results identify complex-field continuity,
rather than endpoint intensity fidelity alone, as the relevant design
principle for stable dynamic holographic control of reconfigurable
neutral-atom arrays.

\textit{A phase-stable criterion for hologram refresh.}---
We consider holographic optical tweezers generated by a phase-only
SLM in a Fourier-plane configuration
~\cite{bergamini2004holographic,DiLeonardo:07}. At the trap centers,
the complex field generated by an SLM phase pattern
\(\boldsymbol{\phi}\) is
\begin{equation}
\label{eq:trap-field}
E_n(\boldsymbol{\phi})
=
\sum_{j=1}^{M}
A_{nj}e^{i\phi_j},
\end{equation}
where \(\phi_j\) is the phase applied to SLM pixel \(j\), and
\(A_{nj}\) is the complex transfer coefficient from pixel \(j\) to
trap \(n\). We denote the trap-field vector in frame \(l\) by
\(\mathbf E^{(l)}
=(E_1^{(l)},\ldots,E_N^{(l)})^T\), with intensity $I_n^{(l)}=|E_n^{(l)}|^2$ and trap phase $\varphi_n^{(l)}=\operatorname{Arg}E_n^{(l)}$.
For a static intensity-only hologram, the endpoint trap depth is
determined by \(I_n^{(l)}\), whereas the associated trap phase
\(\varphi_n^{(l)}\) remains largely unconstrained
~\cite{Gerchberg1972APA,DiLeonardo:07,Pasienski2008,
tamura2016highly,cai2020rapid}.

During a finite SLM refresh from frame \(l\) to frame \(l+1\), this
otherwise unconstrained phase becomes dynamically relevant. The
pixel-level refresh model gives the leading-order transient field at
trap \(n\) as
\begin{equation}
\label{eq:transient_tweezers}
E_{l,n}(a)
\simeq
aE_n^{(l)}
+
(1-a)E_n^{(l+1)},
\end{equation}
where \(a=a(t)\in[0,1]\) is determined by the liquid-crystal response. The pixel-level derivation and higher-order corrections are provided in the Supplemental Material~\cite{supplemental}. The
higher-order terms retain the same phase-sensitive interference
structure. Defining the relative endpoint phase by
$
\Delta\varphi_{l,n}
=
\operatorname{Arg}
\left[
E_n^{(l+1)}E_n^{(l)*}
\right]$,
the interference between the two endpoint fields becomes increasingly
destructive as \(\Delta\varphi_{l,n}\) approaches \(\pi\). For equal endpoint intensities \(I_n^{(l)}=I_n^{(l+1)}\), Eq.~\eqref{eq:transient_tweezers} gives
\begin{equation}
\label{eq:equal-depth-limit}
\min_{a\in[0,1]}
\frac{I_{l,n}(a)}
{I_n^{(l)}}
=
\cos^2
\left(
\frac{\Delta\varphi_{l,n}}{2}
\right),
\end{equation}
where the minimum occurs at \(a=1/2\). 
Thus, two holograms with
identical endpoint intensities can nevertheless exhibit an arbitrarily
deep transient intensity minimum when their endpoint phases differ by
approximately \(\pi\). 
Stable endpoint intensities alone therefore do
not guarantee stable confinement during the refresh.

For a red-detuned tweezer, the optical potential follows the local
intensity, \(U\propto I\), and the trap frequencies scale as
\(\omega_i\propto\sqrt{U}\) in the local Gaussian approximation
~\cite{grimm2000optical}. A sufficiently deep and rapid transient
intensity reduction weakens the instantaneous confinement and can
induce nonadiabatic motional excitation, heating, or loss. More
generally, spatially nonuniform interference may also displace or
distort the local trapping potential.

To quantify the weakest local intensity encountered during each trap
update, we define the normalized transient intensity
\begin{equation}
\label{eq:qnl}
q_{l,n}(a)
=
\frac{I_{l,n}(a)}
{I_n^{(0)}},
\qquad
q_{\min}
=
\min_{l,n}
\min_{a\in[0,1]}
q_{l,n}(a),
\end{equation}
where \(I_n^{(0)}\) is the prescribed initial intensity of trap \(n\).
The trap label follows the corresponding tweezer along its assigned
trajectory, so \(I_n^{(0)}\) provides a fixed reference throughout the
transport sequence. Within the local Gaussian approximation, and provided that the trap
shape and center do not change substantially,
\(U_{l,n}(a)/U_n^{(0)}\simeq q_{l,n}(a)\) and
\(\omega_{i,l,n}(a)/\omega_{i,n}^{(0)}
\simeq \sqrt{q_{l,n}(a)}\) for \(i=r,z\). The low tail of \(q_{l,n}\) therefore identifies the trap updates most
susceptible to transient loss of confinement.

Because \(q_{l,n}(a)\) is determined by both the endpoint amplitudes
and their relative phase, suppressing its low tail requires preserving
the endpoint intensities while maintaining phase continuity between
successive frames. Defining
\(\mathbf I^{(l)}=(I_1^{(l)},\ldots,I_N^{(l)})^T\) and
\(\Delta\boldsymbol{\varphi}^{(l+1)}
=\mathbf{Arg}[\mathbf E^{(l+1)}
\odot\mathbf E^{(l)*}]\), the two corresponding objectives are
\begin{equation}
\label{eq:refresh-criteria}
\min_{\boldsymbol{\phi}^{(l+1)}}
\left\|
\mathbf I^{(l+1)}
\oslash
\mathbf I^{(l+1)}_{\rm tar}
-
\mathbf 1
\right\|_2
\quad \text{and} \quad
\min_{\boldsymbol{\phi}^{(l+1)}}
\left\|
\Delta\boldsymbol{\varphi}^{(l+1)}
\right\|_2,
\end{equation}
where \(\oslash\) and \(\odot\) denote elementwise division and
multiplication, respectively, and \(\mathbf{Arg}\) is evaluated
elementwise. The first objective preserves the prescribed trap depths throughout the transport, whereas the second suppresses inter-frame phase slips that generate destructive transient interference. Together, these objectives define a phase-stable criterion subject to prescribed trap amplitudes.

\begin{figure*}[htbp]
    \centering
    \includegraphics[width=1\textwidth]{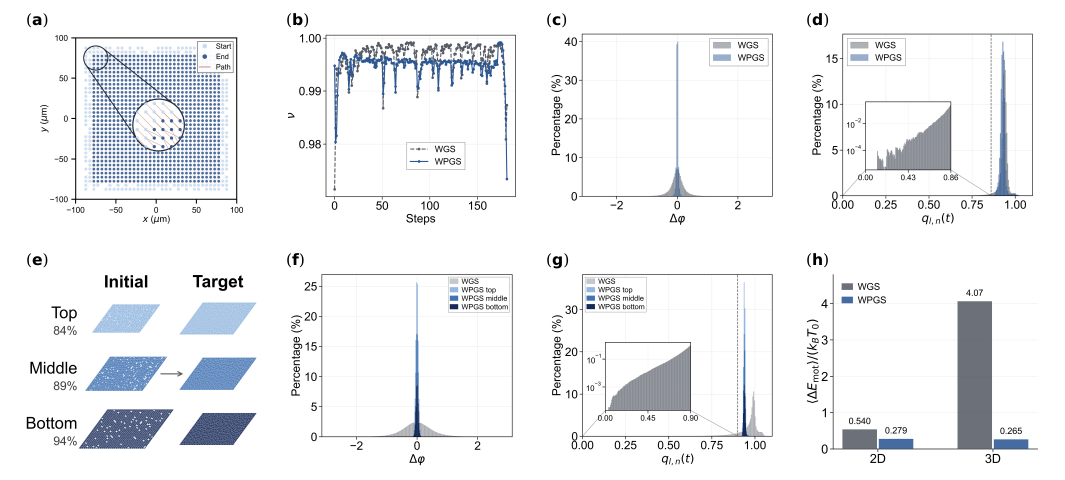}
\caption{ (a)--(d) WGS--WPGS comparison for \(1024\) traps transported from a \(79\%\)-filled \(36\times36\) source array to a \(32\times32\) target array with \(5~\mu\mathrm{m}\) spacing. (a) Transport trajectories, with enlarged representative regions. The mean and maximum transport distances are
\(\overline{\Delta r}=8.24~\mu\mathrm{m}\) and \(\Delta r_{\max}=18.03~\mu\mathrm{m}\).
(b) Endpoint intensity uniformity, \(\nu=1-(I_{\max}-I_{\min})/(I_{\max}+I_{\min})\), along the transport sequence.
(c) Inter-frame phase mismatch \(\Delta\varphi_{l,n}\), aggregated over all traps and transitions. The standard deviation is $0.2381$ for WGS and $0.0291$ for WPGS.
(d) Normalized transient trap intensity \(q_{l,n}(a)\), defined in Eq.~\eqref{eq:qnl}, pooled over all traps, transitions, and sampled refresh points. The inset resolves the WGS low-\(q\) tail.
(e)--(g) Simultaneous 3D reconfiguration of
\(N_{\mathrm{tot}}=3072\) traps at \(z=-30,0,+30~\mu\mathrm{m}\).
(e) Initial and target geometries. The top, middle, and bottom layers start from \(35\times35\), \(34\times34\), and \(33\times33\) source arrays with spacings of \(4\), \(5\), and \(6~\mu\mathrm{m}\), and
filling fractions of \(84\%\), \(89\%\), and \(94\%\), respectively. All layers are transported simultaneously into \(32\times32\) target arrays with \(5~\mu\mathrm{m}\) spacing.
(f) Layer-resolved WPGS phase mismatch distributions; the all-layer WGS distribution is shown in gray.
(g) Layer-resolved WPGS distributions of \(q_{l,n}(a)\); the all-layer WGS distribution is shown in gray. The inset resolves the WGS low-\(q\) tail.
(h) Ensemble-averaged motional heating during transport, \(\langle\Delta E_{\rm mot}\rangle/(k_BT_0)\), for the complete 2D and 3D tasks, computed using the classical trajectory model described in the Supplemental Material~\cite{supplemental}.}
\label{fig:rearrangement}
\end{figure*}

\textit{Implementing the phase-stable criterion.}---
The criterion above specifies the complex-field behavior required for
a stable refresh, whereas the available control variable is only the
phase pattern of the SLM. The implementation problem is therefore to
select, among the fields realizable by a phase-only hologram, an
endpoint field that preserves the prescribed trap intensities while
remaining phase aligned with the preceding frame. For the update \(l\to l+1\), we therefore
combine the two requirements in Eq.~\eqref{eq:refresh-criteria} into
the phase-referenced complex target
\(
\mathbf E_{\rm tar}^{(l+1)}
=
\sqrt{\mathbf I^{(l+1)}_{\rm tar}}
\odot
\exp\!\left(i\boldsymbol{\varphi}^{(l)}\right)
\). To maintain constant trap depths throughout the transport, we set the target intensities to their initial values, $\mathbf I^{(l+1)}_{\rm tar} = I^{(0)}$.

In general, a phase-only SLM cannot reproduce the prescribed complex
field exactly and simultaneously at all trap sites. We therefore seek
the closest realizable field using the weighted complex-field matching
objective
\begin{equation}
\label{eq:optimization-problem}
\min_{\boldsymbol{\phi},s,W}
J(\boldsymbol{\phi},s,W)
:=
\left\|
W\mathbf E(\boldsymbol{\phi})
-
s\mathbf E_{\rm tar}^{(l+1)}
\right\|_2^2 ,
\end{equation}
where \(W=\operatorname{diag}(w_1,\ldots,w_N)\) is a positive
trap-resolved weighting operator and \(s\) is a global complex scale.
For fixed \(\boldsymbol{\phi}\) and \(W\), projection onto the complex
ray spanned by \(\mathbf E_{\rm tar}^{(l+1)}\) gives
\begin{equation}
\label{eq:optimal-scale}
s_\star
=
\frac{
\mathbf E_{\rm tar}^{(l+1)\dagger}
W\mathbf E(\boldsymbol{\phi})
}{
\left\|
\mathbf E_{\rm tar}^{(l+1)}
\right\|_2^2
}.
\end{equation}
Motivated by this nonconvex objective, WPGS provides an efficient
alternating approximation based on amplitude reweighting, complex
projection, and phase-only back-propagation. 
The weighted-projective update seeks a phase-only hologram whose synthesized field closely approximates the phase-referenced target. 
The positive weights \(W\) rebalance trap-dependent amplitude errors, while the complex scale  \(s_\star\) accounts for the common amplitude and global-phase offset between the synthesized and target fields. The phase of  \(s_\star\) fixes the global phase gauge used to evaluate inter-frame phase continuity.
The optimization therefore focuses on the relative complex-field structure across the array and selects a realizable hologram that preserves the prescribed trap intensities while remaining phase aligned with the preceding frame. This suppresses destructive interference during the refresh. Implementation details are provided in the Supplemental Material~\cite{supplemental}.

\textit{Uniform-intensity 2D and 3D
reconfiguration.}---
Large-scale neutral-atom platforms increasingly rely on deterministic
rearrangement to prepare defect-free tweezer arrays, while recent
multiplane control extends this requirement beyond a single plane
~\cite{Schymik2020Enhanced,Pichard2024Rearrangement,
Lin2025AIEnabledRapidAssembly,Kusano2025PlaneSelective}. Under these conditions, dynamic hologram generation must preserve confinement across
many simultaneous trap updates and over long transport sequences. We
therefore compare WGS and WPGS in a \(1024\)-trap 2D reconfiguration
and the simultaneous reconfiguration of \(3072\) traps in three
planes. The corresponding geometries are shown in
Fig.~\ref{fig:rearrangement}(a) and (e), respectively. Both methods use
identical geometries and prescribed trajectories, with source--target
assignments obtained using the Hungarian algorithm
~\cite{kuhn1955hungarian}.

For the 2D reconfiguration, WGS and WPGS maintain comparable endpoint
intensity uniformity throughout the transport, as shown in
Fig.~\ref{fig:rearrangement}(b), demonstrating that enforcing the phase-stable criterion does not compromise endpoint quality.
During the finite refresh, however, WPGS reduces the maximum
frame-to-frame phase mismatch \(\max_{l,n}|\Delta\varphi_{l,n}|\) from \(2.4222\) to
\(0.3801~\mathrm{rad}\), as shown in
Fig.~\ref{fig:rearrangement}(c). Correspondingly, \(q_{\min}\)
increases from \(0.118\) to \(0.866\) for WPGS, as shown in
Fig.~\ref{fig:rearrangement}(d). To assess how the transient intensity suppression affects the
transported atoms, we replay the complete sequence of SLM phase
patterns generated by both methods through the pixel-level SLM refresh
model. This reconstructs the time-dependent spatial optical field and
local trapping potential of every moving tweezer throughout the
transport sequence. Neither method produces transient trap splitting in the 2D task. Accordingly, each transported tweezer remains a single local potential well throughout the refresh, so the normalized intensity \(q_{l,n}(a)\) directly tracks the corresponding variation in trap depth within the local Gaussian approximation. In the corresponding trajectory analysis, WPGS reduces the mean
motional heating during transport \(\langle\Delta E_{\rm mot}\rangle/(k_BT_0)\) from \(0.5396\) to \(0.2789\), as shown in Fig.~\ref{fig:rearrangement}(h).

The 3D reconfiguration is more demanding because traps in different planes follow distinct trajectories while being generated by a common phase-only hologram. Over all traps, layers, and frame transitions, WPGS reduces \(\max_{l,n}|\Delta\varphi_{l,n}|\) from \(2.773~\mathrm{rad}\) to \(0.495~\mathrm{rad}\), as shown in Fig.~\ref{fig:rearrangement}(f). Correspondingly, \(q_{\min}\) increases from \(0.0335\) to \(0.915\), as shown in Fig.~\ref{fig:rearrangement}(g), and the layer-resolved distributions confirm that the improvement persists in all three planes. Applying the same analysis as in the 2D task, we find transient trap splitting for WGS in all three layers, whereas each tweezer for WPGS retains a single local trapping well throughout the transport. The WPGS intensity evolution therefore remains consistent with a single moving trap whose depth varies during the refresh. By contrast, the transient splitting observed for WGS redistributes the optical intensity among competing local maxima, which may increase the likelihood of the atom deviating from the intended trajectory and thereby raise the risk of motional excitation or loss. In the corresponding trajectory analysis, WPGS reduces \(\langle\Delta E_{\rm mot}\rangle/(k_BT_0)\) from \(4.0662\) to \(0.2655\), as shown in Fig.~\ref{fig:rearrangement}(h).

Taken together, the 2D and 3D results show that WPGS preserves endpoint intensity quality while reducing frame-to-frame phase mismatch, raising the transient-intensity floor, and suppressing motional heating during transport. In the more demanding 3D task, it also avoids the transient trap splitting observed for WGS. Similar inter-frame phase continuity and transient-intensity trends are obtained under Gaussian SLM-plane illumination, indicating that these optical results are not restricted to uniform input profiles, as detailed in the Supplemental Material~\cite{supplemental}.

\begin{figure}[htbp]
    \centering
    \includegraphics[width=1\linewidth]{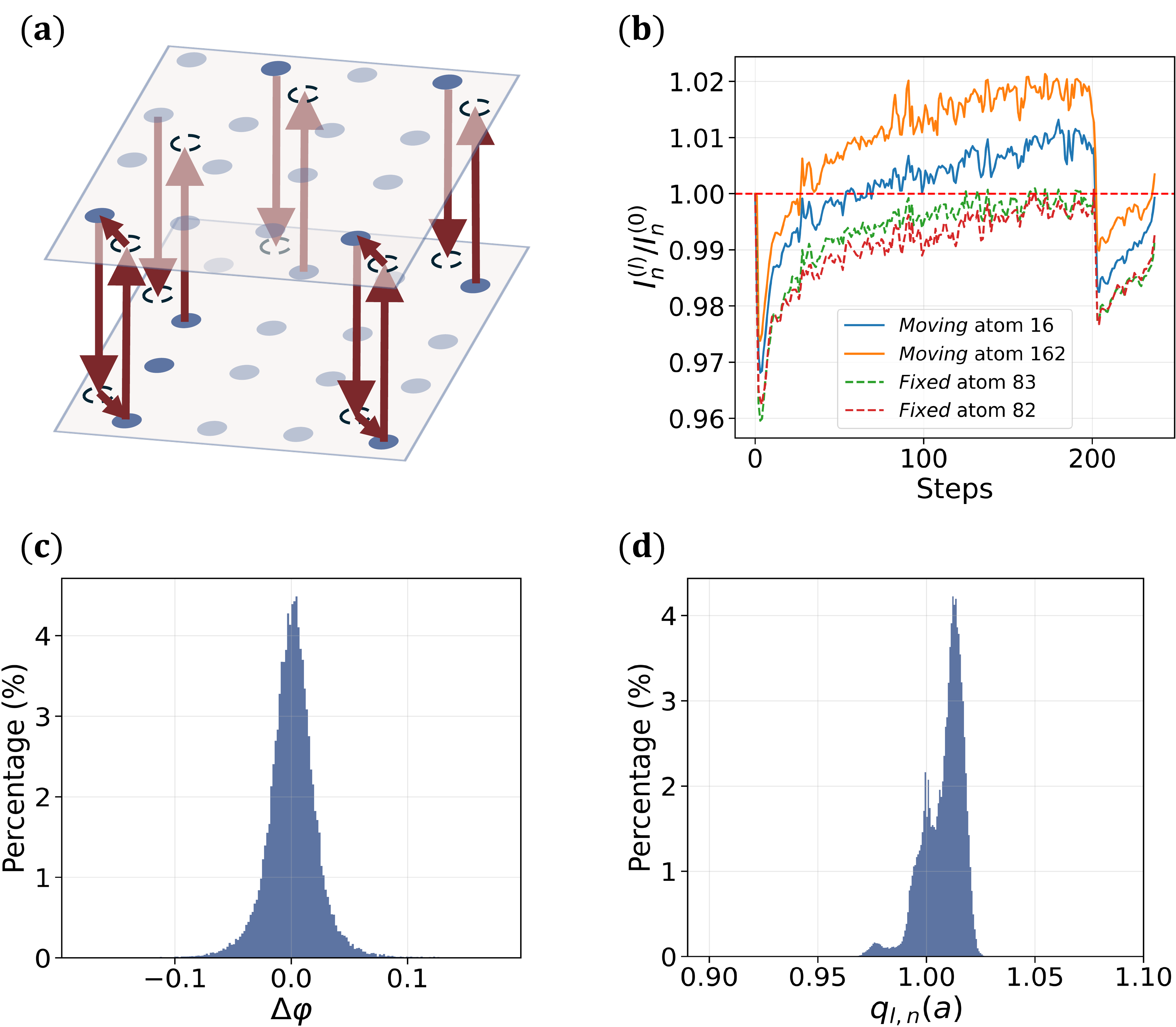}
\caption{\textbf{Phase-stable updates for offset-bilayer transport with
site-dependent target intensities.}
(a) Offset-bilayer geometry in which each trap is assigned a distinct target intensity. Only a subset of traps is transported, including interlayer exchange and additional in-plane motion to fill vacant target sites.
(b) Representative endpoint-intensity ratios \(I_n^{(l)}/I_n^{(0)}\) for moving and stationary traps.
(c) Distribution of the frame-to-frame trap-phase mismatch \(\Delta\varphi_{l,n}\), with standard deviation \(\sigma_{\Delta\varphi}=0.0214~\mathrm{rad}\). 
(d) Corresponding distribution of the normalized transient intensity
\(q_{l,n}(a)\).}
    \label{fig:swapping}
\end{figure}

\textit{Nonuniform-intensity interlayer transport.}---
The phase-referenced complex-field construction permits the target amplitude of each trap to be specified independently while inheriting the phase of the preceding frame. We test whether this local amplitude freedom remains compatible with phase-stable criterion in the offset-bilayer transport shown in Fig.~\ref{fig:swapping}(a).

The representative traces in Fig.~\ref{fig:swapping}(b) show that both moving and stationary traps remain close to their prescribed endpoint intensities. The frame-to-frame trap-phase mismatch is narrowly distributed, with \(\sigma_{\Delta\varphi}=0.0214~\mathrm{rad}\), as shown in Fig.~\ref{fig:swapping}(c). The distribution of \(q_{l,n}(a)\) in Fig.~\ref{fig:swapping}(d) gives \(q_{\min}=0.965\), indicating that the transient-intensity floor remains high despite the site-dependent target amplitudes. The corresponding trajectory analysis gives \(\langle\Delta E_{\rm mot}\rangle/(k_BT_0)=0.140\), with no recapture failures observed among the sampled trajectories; details are provided in Supplemental Material~\cite{supplemental}. These results show that phase-stable hologram updates remain compatible with independently prescribed trap intensities.

\textit{Discussion.}--- Our results show that maintaining inter-frame phase continuity while matching the prescribed trap amplitudes stabilizes transient optical confinement during finite SLM refresh. Holograms with comparable endpoint intensity quality can nevertheless exhibit markedly different behavior during refresh because inter-frame trap-phase mismatch enters the transient field through phase-sensitive interference. Across the uniform-intensity 2D and simultaneous 3D tasks, WPGS reduces the worst-case phase mismatch, raises the transient intensity floor, and lowers motional heating during transport without compromising endpoint quality. In the three-dimensional task, WPGS also avoids the transient trap splitting observed for WGS. Similar inter-frame phase continuity and transient-intensity trends stability are maintained under Gaussian SLM-plane illumination further demonstrate the robust optical performance of WPGS across different input-beam profiles.

The central distinction from WGS-type methods that optimize only the endpoint intensity is that WPGS treats the trap phases as explicit inter-frame constraints, rather than leaving them as unconstrained outputs of the hologram optimization~\cite{DiLeonardo:07,Pasienski2008,kim2019large}. The realized phase of each trap is inherited by the next complex-field target, so endpoint intensity accuracy and phase continuity are addressed within the same hologram update. Existing phase-aware approaches include prescribed phase interpolation between endpoint fields~\cite{Knottnerus2025Parallel} and learned hologram generation based on data-driven training ~\cite{Lin2025AIEnabledRapidAssembly}. WPGS instead determines the phase-only hologram by weighted-projective matching to an inherited complex-field target, operating directly from the optical propagation model without a prescribed interpolation schedule or a data-driven training stage. This model-based formulation provides a direct connection between the optimization objective, inter-frame trap-phase mismatch, and the resulting transient-intensity distribution. In the implementations benchmark, WPGS reaches the prescribed convergence criterion within five iterations and generates each subsequent hologram within a few milliseconds, as detailed in the Supplemental Material~\cite{supplemental}.

Taken together, these results establish the phase-stable criterion as a central design principle for dynamic optical-tweezer control. By explicitly suppressing transient degradation during the refresh, WPGS provides a physically grounded framework for stable holographic reconfiguration at large scale, while remaining compatible with simultaneous multilayer transport and independently prescribed nonuniform trap intensities. The same framework provides a basis for multilayer or heterogeneous
tweezer architectures requiring confinement that varies across layers or atomic species ~\cite{Kusano2025PlaneSelective,Sheng2022DefectFree, Singh2022DualElement,Anand2024DualSpeciesRydberg}. More broadly, the present framework may be extended to jointly optimize transport trajectories, trap-dependent confinement, and the full refresh evolution.

\textbf{\textit{Acknowledgements.}}---The authors would like to thank Jingbo Wang and Weijia Wen for helpful comments. This work was partially supported by the National Natural Science Foundation of China (Grant Nos. 92576114 and 12404568) and the Guangdong Provincial Quantum Science Strategic Initiative (Grant Nos. GDZX2403008 and GDZX2503001).

%

\onecolumngrid

\appendix
\setcounter{subsection}{0}
\setcounter{table}{0}
\setcounter{figure}{0}

\vspace{3cm}

\begin{center}
\Large{\textbf{Supplemental Material} \\ \textbf{
}}
\end{center}

\renewcommand{\theequation}{S\arabic{equation}}
\renewcommand{\theproposition}{S\arabic{proposition}}
\renewcommand{\thedefinition}{S\arabic{definition}}
\renewcommand{\thefigure}{S\arabic{figure}}
\setcounter{equation}{0}
\setcounter{table}{0}
\setcounter{section}{0}
\setcounter{proposition}{0}
\setcounter{definition}{0}
\setcounter{figure}{0}

\section{Optical propagation model}
\label{sec:sm-propagation}

 We consider an optical system in which a SLM occupies the front focal plane of a thin lens with focal length $f$. Optical traps form in the back focal region at positions $\{\mathbf{r}_n\}_{n=1}^N$, where $\mathbf{r}_n = (x_n, y_n, z_n)$ denotes the center of the $n$-th trap. The SLM comprises $M = M_x \times M_y$ pixels arranged on a rectangular grid with pitch $d$; pixel $j$ is centered at transverse coordinates $(x_j, y_j)$. Under uniform illumination with amplitude $u_0$, the field reflected from pixel $j$ is $u_0 e^{i\phi_j}$, where $\phi_j \in [0, 2\pi)$ is the programmable phase. In the Fresnel approximation, the optical field at trap $n$ is obtained by summing contributions from all pixels~\cite{DiLeonardo:07}, 
\begin{equation}
E_{n}
= \frac{e^{i 2\pi (f+z_n)/\lambda}}{i\lambda(f+z_n)}\, d^2 u_{0} 
\sum_{j=1}^M \exp\!\bigl[i\bigl(\phi_{j} + \Delta_j^n\bigr)\bigr],
\end{equation}
where the phase offset is written in the same axial-coordinate convention as the separable implementation below,
\begin{equation}\label{eq:Delta}
    \Delta_j^n = -\frac{\pi z_n}{\lambda f^2}(x_j^2 + y_j^2)
      - \frac{2\pi}{\lambda f}(x_n x_j + y_n y_j).
\end{equation}
The complex field at the trap centers generated by an SLM phase pattern $\boldsymbol{\phi} = \{\boldsymbol{\phi}_j\}_{j=1}^M$ can therefore be written as
\begin{equation}
E_n(\boldsymbol{\phi})
=
\sum_{j=1}^{M} A_{nj}e^{i\phi_j},
\end{equation}
where $\phi_j$ is the phase applied by SLM pixel $j$, and  the propagation matrix $A \in \mathbb{C}^{N \times M}$ introduced in the main text has entries
\begin{equation}\label{eq:A-matrix}
    A_{nj} = \frac{d^2 u_0}{\lambda f} \cdot
      \frac{e^{i 2\pi (2f + z_n)/\lambda}}{i} \, e^{i \Delta_j^n},
\end{equation}
where $\lambda$ is the illumination wavelength.

\section{Optical-field model during SLM refresh}
\label{sec:sm-refresh}

\subsection{Exact transient field within the pixel-relaxation model}

 During an SLM refresh, the liquid-crystal response at each pixel is well approximated by exponential relaxation~\cite{Lin2025AIEnabledRapidAssembly}:
\begin{equation}\label{eq:phase-transition}
    \phi_j(t)
    =
    a(t)\,\phi_j^{(l)}
    +
    [1-a(t)]\,\phi_j^{(l+1)},
\end{equation}
where $a(t)=e^{-(t-t_0)/\tau}$ decays from unity at the start of the transition ($t=t_0$) toward zero with time constant $\tau$. Here superscripts $(l)$ and $(l+1)$ denote the initial and final hologram frames. Defining the per-pixel phase excursion $\Delta\phi_j := \mathrm{wrap}(\phi_j^{(l+1)}-\phi_j^{(l)})$ and substituting Eq.~\eqref{eq:phase-transition} into $E_n=\sum_{j=1}^{M}A_{nj}e^{i\phi_j}$ gives
\begin{equation}\label{eq:En-t}
    E_n(t)
    =
    \sum_{j=1}^{M}
    A_{nj}\,
    e^{i[\phi_j^{(l)}+(1-a)\Delta\phi_j]} .
\end{equation}

To separate the contributions from the initial and final holograms, we use
\begin{equation}
e^{i[\phi_j^{(l)}+(1-a)\Delta\phi_j]}
=
e^{i\phi_j^{(l)}}
\frac{\sin(a\Delta\phi_j)}{\sin(\Delta\phi_j)}
+
e^{i\phi_j^{(l+1)}}
\frac{\sin[(1-a)\Delta\phi_j]}{\sin(\Delta\phi_j)}.
\end{equation}
Substituting this identity into Eq.~\eqref{eq:En-t} yields the exact transient field~\cite{Lin2025AIEnabledRapidAssembly}:
\begin{equation}\label{eq:transient-field-exact}
    E_n(t)
    =
    \sum_{j=1}^{M}
    A_{nj}
    \left[
        e^{i\phi_j^{(l)}}
        \frac{\sin(a\Delta\phi_j)}{\sin(\Delta\phi_j)}
        +
        e^{i\phi_j^{(l+1)}}
        \frac{\sin[(1-a)\Delta\phi_j]}{\sin(\Delta\phi_j)}
    \right].
\end{equation}

\subsection{Leading-order approximation}

When the inter-frame phase excursion is small, Eq.~\eqref{eq:transient-field-exact} admits a simple leading-order approximation~\cite{Lin2025AIEnabledRapidAssembly}. Using $\sin(ax)/\sin x=a+O(x^2)$ and $\sin[(1-a)x]/\sin x=(1-a)+O(x^2)$ for $|x|\ll1$, we obtain
\begin{equation}\label{eq:transient-field-leading}
    \hat{{\mathbf{E}}}^{(l+1)}(t)
    \approx
    a(t)\,\mathbf{E}^{(l)}
    +
    [1-a(t)]\,\mathbf{E}^{(l+1)},
\end{equation}
where $E_n^{(q)}=\sum_{j=1}^{M}A_{nj}e^{i\phi_j^{(q)}}$ for $q\in\{l,l+1\}$. Equation~\eqref{eq:transient-field-leading} is the leading-order form used in the main text. It shows that, for small phase excursions, the transient field is approximately an interpolation between the initial and final trap fields, weighted by the pixel-relaxation factor $a(t)$.

\subsection{Higher-order correction}
A more accurate approximation is obtained by retaining the $O(\Delta\phi_j^2)$ terms in the ratio expansions:
\begin{align}
    \frac{\sin(a\,\Delta\phi_j)}{\sin(\Delta\phi_j)}
    &=
    a\left[
        1+\frac{(1-a^2)\Delta\phi_j^2}{6}
    \right]
    +O(\Delta\phi_j^4), \label{eq:ratio-a} \\
    \frac{\sin[(1-a)\Delta\phi_j]}{\sin(\Delta\phi_j)}
    &=
    (1-a)\left[
        1+\frac{a(2-a)\Delta\phi_j^2}{6}
    \right]
    +O(\Delta\phi_j^4). \label{eq:ratio-1-a}
\end{align}
Substituting Eqs.~\eqref{eq:ratio-a} and \eqref{eq:ratio-1-a} into Eq.~\eqref{eq:transient-field-exact} yields
\begin{equation}
E_n(t)
\approx
a\sum_{j=1}^{M}A_{nj}e^{i\phi_j^{(l)}}
\left[
    1+\frac{(1-a^2)\Delta\phi_j^2}{6}
\right]
+
(1-a)\sum_{j=1}^{M}A_{nj}e^{i\phi_j^{(l+1)}}
\left[
    1+\frac{a(2-a)\Delta\phi_j^2}{6}
\right].
\end{equation}
We decompose $\Delta\phi_j^2$ into its mean and fluctuation parts:
\begin{equation}
\Delta\phi_j^2=\langle \Delta\phi^2\rangle+\varepsilon_j,
\qquad
\langle \Delta\phi^2\rangle:=\frac{1}{M}\sum_{j=1}^{M}\Delta\phi_j^2,
\qquad
\sum_{j=1}^{M}\varepsilon_j=0.
\end{equation}
Then, for $q\in\{l,l+1\}$,
\begin{equation}
\sum_{j=1}^{M}A_{nj}e^{i\phi_j^{(q)}}\Delta\phi_j^2
=
\langle \Delta\phi^2\rangle E_n^{(q)} + R_n^{(q)},
\qquad
R_n^{(q)}
:=
\sum_{j=1}^{M}A_{nj}e^{i\phi_j^{(q)}}\varepsilon_j.
\end{equation}
By the Cauchy--Schwarz inequality,
\begin{equation}\label{eq:R-bound}
    |R_n^{(q)}|
    \le
    \left(\sum_{j=1}^{M}|A_{nj}|^2\right)^{1/2}
    \left(\sum_{j=1}^{M}\varepsilon_j^2\right)^{1/2}.
\end{equation}
Hence, whenever $|R_n^{(q)}|\ll |\langle \Delta\phi^2\rangle E_n^{(q)}|$, the weighted second-moment sum may be approximated by $\langle \Delta\phi^2\rangle E_n^{(q)}$. The field then reduces to
\begin{equation}\label{eq:transient-field-approx}
    \hat{\mathbf{E}}^{(l+1)}(t)
    \approx
    a(t)\,\alpha^{(l)}\,\mathbf{E}^{(l)}
    +
    [1-a(t)]\,\alpha^{(l+1)}\,\mathbf{E}^{(l+1)},
\end{equation}
with
\begin{equation}
\alpha^{(l)}
=
1+\frac{1-a^2}{6}\langle\Delta\phi^2\rangle,
\qquad
\alpha^{(l+1)}
=
1+\frac{a(2-a)}{6}\langle\Delta\phi^2\rangle.
\end{equation}
Under the stated small-fluctuation condition, the leading higher-order correction mainly renormalizes the amplitudes of the two interpolating terms. The destructive-interference mechanism remains controlled by the relative trap phase. Since $\langle\Delta\phi^2\rangle\ll1$, one has $\alpha^{(l)},\alpha^{(l+1)}=1+O(\langle\Delta\phi^2\rangle)$, and Eq.~\eqref{eq:transient-field-leading} is recovered at leading order.

\subsection{Transient intensity and interference term}

Using the leading-order approximation in Eq.~\eqref{eq:transient-field-leading}, the transient intensity at trap $n$ is
\begin{equation}\label{eq:In-t}
    \hat{I}_{l,n}(a)
    =
    |\hat{E}_{l,n}(a)|^2
    \approx
    \bigl|
        a\,E_n^{(l)}
        +
        (1-a)\,E_n^{(l+1)}
    \bigr|^2.
\end{equation}
Writing $E_n^{(q)}=\sqrt{I_n^{(q)}}\,e^{i\varphi_n^{(q)}}$ for $q\in\{l,l+1\}$, we obtain
\begin{equation}\label{eq:In-t-expanded}
\begin{aligned}
    \hat{I}_{l,n}(a)
    &\approx
    a^2 I_n^{(l)}
    +
    (1-a)^2 I_n^{(l+1)} \\
    &\quad +
    2a(1-a)\sqrt{I_n^{(l)}I_n^{(l+1)}}
    \cos\Delta\varphi_{l,n}.
\end{aligned}
\end{equation}
 Thus the transient intensity contains contributions from the initial and final trap intensities together with an interference term controlled by the relative phase between consecutive holograms. When $\Delta\varphi_{l,n}\approx \pi$, the interference term is maximally negative, leading to a significant intensity dip even if $I_n^{(l)}\approx I_n^{(l+1)}$. If the higher-order correction in Eq.~\eqref{eq:transient-field-approx} is retained, the corresponding intensity expression becomes
\begin{equation}\label{eq:In-t-expanded-second-order}
\begin{aligned}
    \hat{I}_{l,n}(a)
    &\approx
    a^2(\alpha^{(l)})^2 I_n^{(l)}
    +
    (1-a)^2(\alpha^{(l+1)})^2 I_n^{(l+1)} \\
    &\quad +
    2a(1-a)\alpha^{(l)}\alpha^{(l+1)}
    \sqrt{I_n^{(l)}I_n^{(l+1)}}
    \cos\Delta\varphi_{l,n}.
\end{aligned}
\end{equation}
Equation~\eqref{eq:In-t-expanded-second-order} shows that the higher-order correction rescales the amplitudes of the three contributions, while the interference mechanism itself remains controlled by the relative phase.

\section{Weighted-projective
Gerchberg–Saxton}
\label{sec:wpgs-details}

To implement the phase-stable criterion defined in the main text, WPGS promotes inter-frame phase continuity by combining the prescribed next-frame amplitudes with the realized phases of the corresponding traps in the preceding frame. To synthesize this phase-referenced target with a phase-only SLM, we formulate a weighted complex-field matching problem. The nonconvex optimization problem is
\begin{equation}\label{eq:optimization-problem-app}
    \min_{\boldsymbol{\phi},\, s,\, W>0}\;
    J(\boldsymbol{\phi}, s, W)
    :=
    \bigl\| W \mathbf{E}(\boldsymbol{\phi}) - s\, \mathbf{E}_{\mathrm{tar}} \bigr\|_2^2,
\end{equation}
where $W=\mathrm{diag}(w_1,\dots,w_N)$ is a positive diagonal matrix,
$s\in\mathbb{C}$ is a global complex scale factor, and
$\mathbf{E}(\boldsymbol{\phi})=A e^{i\boldsymbol{\phi}}$ is the synthesized trap field.
Motivated by Eq.~\eqref{eq:optimization-problem-app}, we construct an
alternating update over the three variable blocks
$(W,s,\boldsymbol{\phi})$. 

\paragraph{Weight update.}
Given $\boldsymbol{\phi}^{(k)}$ and $s^{(k)}$, the weight block is updated from
\begin{equation}\label{eq:w-subproblem-app}
    W^{(k+1)}
    \in
    \arg\min_{W>0}
    \bigl\|
        W \mathbf{E}^{(k)}
        -
        s^{(k)} \mathbf{E}_{\mathrm{tar}}
    \bigr\|_2^2,
\end{equation}
where $\mathbf{E}^{(k)} := \mathbf{E}(\boldsymbol{\phi}^{(k)})$.
Because $W=\mathrm{diag}(w_1,\dots,w_N)$ is diagonal and real-valued,
Eq.~\eqref{eq:w-subproblem-app} decouples into independent scalar problems,
\begin{equation}\label{eq:w-subproblem-scalar-app}
    \min_{w_n>0}\;
    \bigl|
        w_n E_n^{(k)}
        -
        s^{(k)} E_{\mathrm{tar},n}
    \bigr|^2,
    \qquad n=1,\dots,N.
\end{equation}
Since the role of $W$ is specifically to compensate trap-to-trap \emph{amplitude}
nonuniformity, while phase alignment is handled by the scale $s$ and the SLM
phase $\boldsymbol{\phi}$, we replace the phase-sensitive scalar problem in
Eq.~\eqref{eq:w-subproblem-scalar-app} by the phase-insensitive amplitude surrogate
\begin{equation}\label{eq:w-subproblem-amp-app}
    \min_{w_n>0}\;
    \bigl|
        w_n |E_n^{(k)}|
        -
        |s^{(k)}|\,|E_{\mathrm{tar},n}|
    \bigr|^2,
    \qquad n=1,\dots,N.
\end{equation}
This is a one-dimensional convex quadratic in the real variable $w_n$, whose
unique minimizer is
\begin{equation}\label{eq:w-solution-app}
    w_{n,\star}^{(k+1)}
    =
    |s^{(k)}|
    \frac{|E_{\mathrm{tar},n}|}{|E_n^{(k)}|}.
\end{equation}
Because the factor $|s^{(k)}|$ is common to all traps, it cancels in the
subsequent normalization and does not affect the relative reweighting.
We use the amplitude ratio in Eq.~\eqref{eq:w-solution-app} as a
multiplicative correction to the accumulated trap weight. Because the common
factor $|s^{(k)}|$ cancels under the subsequent normalization, this gives
\begin{equation}\label{eq:w-update-raw-app}
    \widehat{w}_n^{(k+1)}
    =
    w_n^{(k)}
    \frac{|E_{\mathrm{tar},n}|}{|E_n^{(k)}|}.
\end{equation}
This multiplicative form preserves positivity and is equivalent to an additive
update in the log-domain. For a uniform target amplitude, $|E_{\mathrm{tar},n}|=\bar{E}_{\mathrm{tar}}$
for all $n$, Eq.~\eqref{eq:w-update-raw-app} reduces to
\begin{equation}\label{eq:w-update-uniform-app}
    \widehat{w}_n^{(k+1)}
    =
    w_n^{(k)}
    \frac{\bar{E}_{\mathrm{tar}}}{|E_n^{(k)}|},
\end{equation}
where $\bar{E}_{\mathrm{tar}} := \frac{1}{N}\sum_{\ell=1}^{N}|E_{\mathrm{tar},\ell}|$.
Because Eq.~\eqref{eq:w-update-raw-app} uses the individual target amplitudes
$|E_{\mathrm{tar},n}|$, the same update extends directly to nonuniform target
intensity patterns. The resulting amplitude-equalization rule is analogous to the
update used in weighted Gerchberg--Saxton schemes~\cite{DiLeonardo:07,nogrette2014single}:
it suppresses over-bright traps and boosts under-bright ones while preserving positivity. We then normalize the updated weights by their arithmetic mean,
\begin{equation}\label{eq:w-update-normalized-app}
    \overline{\widehat{w}}^{(k+1)}
    :=
    \frac{1}{N}\sum_{\ell=1}^N \widehat{w}_{\ell}^{(k+1)},
    \qquad
    \widetilde{w}_n^{(k+1)}
    =
    \frac{\widehat{w}_n^{(k+1)}}{\overline{\widehat{w}}^{(k+1)}}.
\end{equation}
This normalization preserves the relative reweighting, fixes the otherwise
arbitrary common scale shared by $W$ and $s$, and prevents finite-precision
weight drift. Moreover, the common
positive factor $\overline{\widehat{w}}^{(k+1)}$ does not affect the phase
update in Eq.~\eqref{eq:phi-update-app}, since it can be absorbed as a global
amplitude factor before phase extraction. In transport calculations, we optionally apply a mild late-stage relaxation
to the normalized update when the number of tweezers is large,
\begin{equation}\label{eq:w-overrelax-app}
    w_n^{(k+1)}
    =
    w_n^{(k)}
    +
    \beta\bigl(
        \widetilde{w}_n^{(k+1)}
        -
        w_n^{(k)}
    \bigr),
    \qquad
    \beta=0.8.
\end{equation}
This relaxation is used only near the end of transport. Since both $w^{(k)}$ and $\widetilde{w}^{(k+1)}$ are normalized to unit mean, this refinement preserves
the overall weight scale. When this refinement is not used, we simply set
$w_n^{(k+1)}=\widetilde{w}_n^{(k+1)}$.

\paragraph{Scale update.}
Given $\boldsymbol{\phi}^{(k)}$ and $W^{(k+1)}$, the subproblem over $s$ is
\begin{equation}\label{eq:s-subproblem-app}
    s^{(k+1)}
    \in
    \arg\min_{s \in \mathbb{C}}
    \bigl\|
        W^{(k+1)} \mathbf{E}^{(k)}
        -
        s\,\mathbf{E}_{\mathrm{tar}}
    \bigr\|_2^2.
\end{equation}
This is a convex least-squares problem in a single complex scalar. Expanding the
objective gives
\begin{equation}\label{eq:s-expand-app}
\begin{aligned}
    \bigl\|
        W^{(k+1)} \mathbf{E}^{(k)}
        -
        s\,\mathbf{E}_{\mathrm{tar}}
    \bigr\|_2^2
    &=
    \bigl\|
        W^{(k+1)} \mathbf{E}^{(k)}
    \bigr\|_2^2 \\
    &\quad
    -2\,\operatorname{Re}\!\Bigl(
        s^*\,
        \mathbf{E}_{\mathrm{tar}}^\dagger
        W^{(k+1)} \mathbf{E}^{(k)}
    \Bigr)
    +|s|^2\|\mathbf{E}_{\mathrm{tar}}\|_2^2.
\end{aligned}
\end{equation}
Differentiating with respect to $s^*$ and setting the derivative to zero yields
\begin{equation}\label{eq:s-update-app}
    s^{(k+1)}
    =
    \frac{
        \mathbf{E}_{\mathrm{tar}}^\dagger
        \bigl(
            W^{(k+1)} \mathbf{E}^{(k)}
        \bigr)
    }{
        \|\mathbf{E}_{\mathrm{tar}}\|_2^2
    }.
\end{equation}
Thus, the scale step is solved exactly at each iteration by projection of the
current weighted field onto the target field in the complex least-squares sense.
Substituting the optimal scale back into the objective gives
\begin{equation}\label{eq:projective-objective-app}
    J(\boldsymbol{\phi}, s_\star, W)
    =
    \bigl\|
        (I-P_{\mathrm{tar}})(W\mathbf{E})
    \bigr\|_2^2,
\end{equation}
where
\begin{equation}\label{eq:projector-app}
    P_{\mathrm{tar}}
    :=
    \frac{
        \mathbf{E}_{\mathrm{tar}}
        \mathbf{E}_{\mathrm{tar}}^\dagger
    }{
        \|\mathbf{E}_{\mathrm{tar}}\|_2^2
    }
\end{equation}
is the orthogonal projector onto $\mathrm{span}\{\mathbf{E}_{\mathrm{tar}}\}$.
Hence the objective measures the component of the weighted field $W\mathbf{E}$
orthogonal to the target direction, which motivates the term \emph{projective}.

\paragraph{SLM Phase update.}
Given $W^{(k+1)}$ and $s^{(k+1)}$, the subproblem over $\boldsymbol{\phi}$ is
\begin{equation}\label{eq:phi-subproblem-app}
    e^{i\boldsymbol{\phi}^{(k+1)}}
    \in
    \arg\min_{\mathbf{u}:\,|u_j|=1\ \forall j}
    \bigl\|
        W^{(k+1)} (A\mathbf{u})
        -
        s^{(k+1)}\mathbf{E}_{\mathrm{tar}}
    \bigr\|_2^2.
\end{equation}
Unlike the scale step, this problem is nonconvex because of the unit-modulus
constraint on every SLM pixel. In general,
Eq.~\eqref{eq:phi-subproblem-app} does not admit a simple closed-form minimizer.
Projecting this phase-referenced alignment direction onto the unit-modulus SLM constraint yields a phase update with the algebraic form of a Gerchberg--Saxton back-projection~\cite{Gerchberg1972APA,DiLeonardo:07,fienup1982phase}:
we back-propagate the weighted target field through the adjoint operator and
then project onto the unit-modulus constraint by phase extraction,
\begin{equation}\label{eq:phi-update-app}
    \boldsymbol{\phi}^{(k+1)}
    =
    \arg\!\Bigl(
        A^\dagger
        \bigl(
            W^{(k+1)} s^{(k+1)} \mathbf{E}_{\mathrm{tar}}
        \bigr)
    \Bigr),
\end{equation}
where $\arg(\cdot)$ acts componentwise and returns the phase of each complex
entry. This phase step should therefore be understood as an approximate
projection rather than an exact minimizer of
Eq.~\eqref{eq:phi-subproblem-app}; accordingly, it does not in general guarantee
monotone descent of the original objective
$J(\boldsymbol{\phi},s,W)$.

\section{Minimal two-frame refresh mechanism}
\label{sec:sm-minimal-refresh}

\begin{figure}[htbp]
    \centering
    \includegraphics[width=1\linewidth]{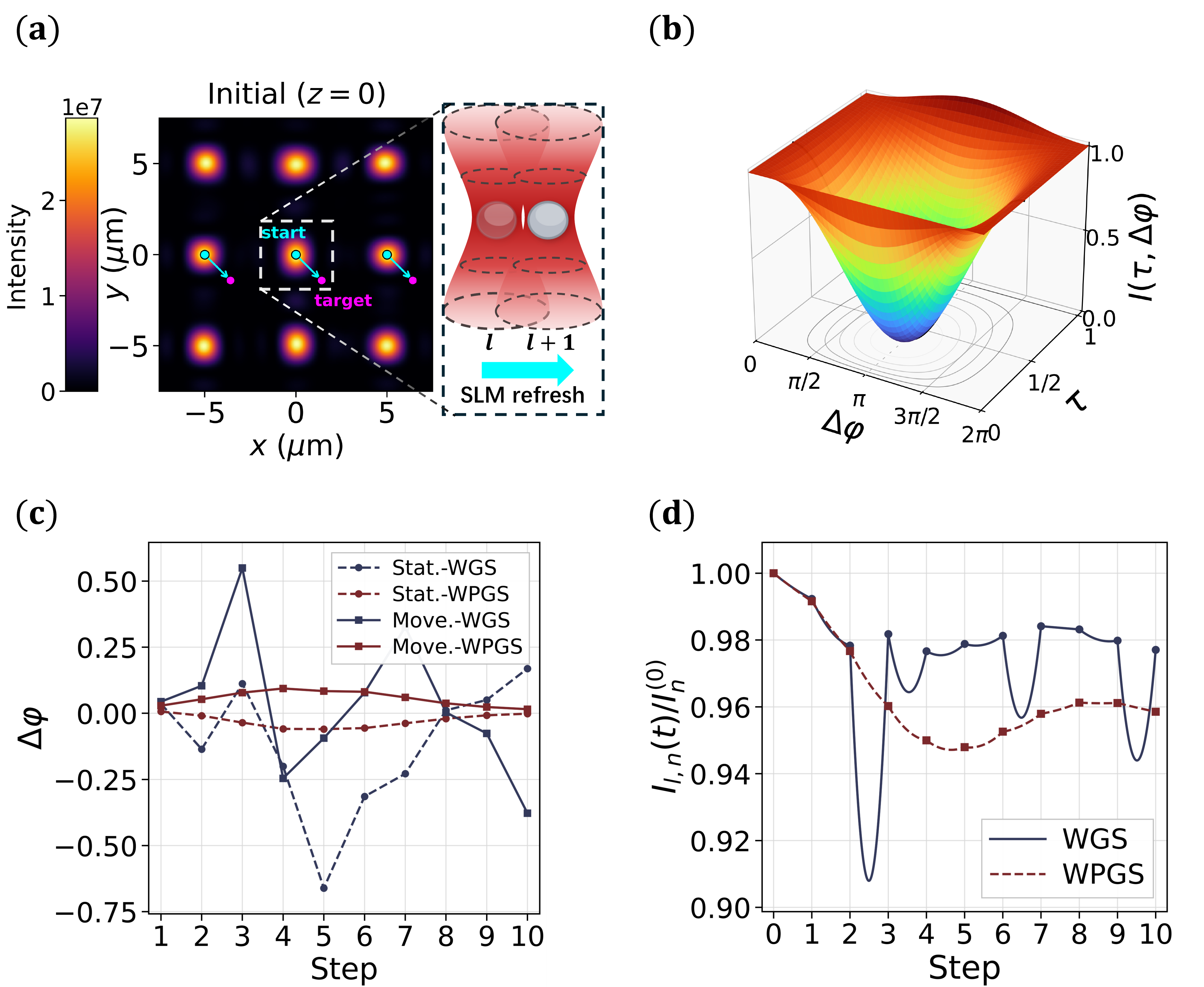}
    \caption{\textbf{Minimal refresh mechanism}
(a) Minimal $3\times3$ transport configuration. The middle row is displaced along the lower-right diagonal, and the inset schematically shows the finite SLM refresh between frames $l$ and $l+1$. 
(b) Normalized equal-depth transient-intensity landscape,
$I(a,\Delta\varphi)=|a+(1-a)e^{i\Delta\varphi}|^2$,
evaluated over the refresh coordinate $a\in[0,1]$ and one period of the inter-frame phase mismatch.
(c) Inter-frame phase mismatch for the representative stationary and moving traps under WGS and WPGS.
(d) Initial-intensity-normalized evolution, $I_{l,n}(t)/I_n^{(0)}$, of the central moving trap over ten consecutive refresh events. The leading-order two-frame model reveals pronounced within-refresh minima under WGS, whereas WPGS suppresses these excursions and maintains a higher minimum intensity.}
    \label{fig:sm-minimal-refresh}
\end{figure}

To provide a minimal numerical test of the frame-pair refresh mechanism, we consider nine traps initially arranged as a \(3\times3\) square array in the \(z=0\) plane, with an in-plane spacing of \(5~\mu\mathrm{m}\) and uniform prescribed trap intensities shown in Fig.~\ref{fig:sm-minimal-refresh}(a). The upper and lower rows remain stationary, while all three traps in the middle row are translated together along the \(-45^\circ\) diagonal. The total displacement is \(2.0~\mu\mathrm{m}\), discretized into ten equal steps of \(0.2~\mu\mathrm{m}\), giving eleven endpoint configurations and ten frame-to-frame refresh events. WGS and WPGS are evaluated using the same prescribed trajectory and optical
propagation model, starting from the same frame-0 hologram and realized trap
field.

For each pair of consecutive endpoint holograms, we evaluate the transient field using the leading-order two-endpoint model in Eq.~\eqref{eq:transient-field-leading}. Within this model, the transient trap intensity depends on the inter-frame phase mismatch \(\Delta\varphi\) as well as the two endpoint depths. Fig.~\ref{fig:sm-minimal-refresh}(b) shows the corresponding normalized equal-depth landscape. The transient dip becomes deeper as \(\Delta\varphi\) approaches \(\pi\), and the intensity vanishes at \(a=1/2\) for equal endpoint amplitudes with \(\Delta\varphi=\pi\). 

Fig.~\ref{fig:sm-minimal-refresh}(c,d) illustrates how the update rule modifies this mechanism in the minimal sequence. WGS produces larger trap-resolved phase mismatch for both the representative moving trap and the nominally stationary trap. WPGS keeps the corresponding phase mismatch closer to zero, consistent with the frame-pair diagnostics. In Fig.~\ref{fig:sm-minimal-refresh}(d), WPGS reduces the within-refresh
excursions of \(I_{l,n}(t)/I_n^{(0)}\) and raises its minimum value. The minimal example therefore illustrates the finite-refresh mechanism
summarized in Fig. 1 of the main text and reproduces the equal-depth
limit \(I_{\min}/I^{(0)}=\cos^2(\Delta\varphi/2)\).

\section{Simulation settings and benchmark tasks}
\label{sec:sm-task-definitions}

This section summarizes the common simulation settings and task
parameters used for Figs. 2 and 3 of the main text. All figures are generated using the same phase-only SLM model introduced in the main text. We use an SLM with $1024\times1024$ pixels and pixel pitch $17~\mu\mathrm{m}$, with hologram-propagation wavelength $\lambda=820~\mathrm{nm}$ and focal length $f=4~\mathrm{mm}$. These parameters define the SLM propagation model used for all hologram
calculations. The atom-dynamics diagnostic below uses a separate nominal Gaussian-trap parameter set.

The two-dimensional and three-dimensional benchmarks in Fig. 2 of the
main text use uniform target amplitudes, whereas the nonuniform-intensity interlayer benchmark in Fig. 3 uses trap-dependent target amplitudes. In all three
tasks, the transport trajectories are discretized so that the maximum displacement per update step is $0.1~\mu\mathrm{m}$.

Panels (a)--(d) of Fig. 2 in the main text show a two-dimensional reconfiguration task with $1024$ target traps. The source is a partially filled $36\times36$ array ($79\%$ filling), and the target is a $32\times32$ array with spacing $5~\mu\mathrm{m}$. The source-to-target assignment is determined using the Hungarian algorithm, which defines the transport trajectories used in the simulation. The mean transport distance is $\overline{\Delta r}=8.24~\mu\mathrm{m}$, and the maximum transport distance is $\Delta r_{\max}=18.03~\mu\mathrm{m}$. 

Panels (e)--(h) of Fig. 2 in the main text show a simultaneous
three-dimensional reconfiguration task with $N_{\mathrm{tot}}=3072$ target traps, distributed over three planes at $z=-30,0,+30~\mu\mathrm{m}$, with $1024$ target traps per layer. The initial layers are nonuniform and are given by a $33\times33$ array with spacing $6~\mu\mathrm{m}$ ($94\%$ filling), a $34\times34$ array with spacing $5~\mu\mathrm{m}$ ($89\%$ filling), and a $35\times35$ array with spacing $4~\mu\mathrm{m}$ ($84\%$ filling), respectively. The target consists of identical $32\times32$ arrays with spacing $5~\mu\mathrm{m}$ in all three layers. The source-to-target assignment is likewise determined using the Hungarian algorithm within each layer.

Figure 3 of the main text considers a nonuniform-intensity interlayer transport task composed of two $10\times10$ arrays. The in-plane spacing is $5~\mu\mathrm{m}$, the axial separation is $20~\mu\mathrm{m}$, and the two layers are laterally offset by $2.5~\mu\mathrm{m}$. The target is nonuniform, i.e., the target amplitude is trap dependent.


\section{Robustness to Gaussian SLM-plane illumination}
\label{sec:sm-gaussian-illumination}

To test whether the refresh-stability advantage of WPGS depends on the
assumed SLM-plane illumination profile, we repeated the complete hologram
calculations using a centered Gaussian input amplitude, as commonly adopted
for input-plane fields in iterative hologram design and holographic optical
tweezers~\cite{Pasienski2008}
\begin{equation}
A_R(x,y)
=
C_R\exp\!\left[-\frac{x^2+y^2}{R^2}\right],
\qquad x,y\in[-1,1],
\label{eq:sm-gaussian-input}
\end{equation}
where \(x\) and \(y\) are expressed in units of the SLM half-width. The
normalization constant \(C_R\) is chosen such that
\(\langle |A_R|^2\rangle_{\rm SLM}=1\), keeping the mean incident intensity
fixed. We consider \(R=0.5,1.0,\) and \(1.5\), representing concentrated,
intermediate, and broad illumination profiles over the active SLM area. For
each value of \(R\), the same input amplitude is used consistently in the
reference-hologram optimization, all sequential hologram updates, and the
evaluation of the refresh-resolved field.

\begin{table}[htbp]
\caption{\textbf{Refresh stability under Gaussian SLM-plane illumination.}
The Gaussian radius \(R\) is expressed in units of the SLM half-width.
Each entry gives the minimum normalized transient intensity \(q_{\min}\) over all traps and
frame-to-frame refresh events in the corresponding transport sequence.}
\centering
\label{tab:sm-gaussian-illumination}
\begin{ruledtabular}
\begin{tabular}{llrrr}
Task & Method &
\(R=0.5\) &
\(R=1.0\) &
\(R=1.5\) \\
\hline
2D reconfiguration & WGS
    & $0.0959$ & $0.3042$ & $0.1820$ \\
2D reconfiguration & WPGS
    & $\mathbf{0.8806}$ & $\mathbf{0.8724}$ & $\mathbf{0.8766}$ \\
3D reconfiguration & WGS
    & $0.0474$ & $0.0490$ & $0.0391$ \\
3D reconfiguration & WPGS
    & $\mathbf{0.8957}$ & $\mathbf{0.9156}$ & $\mathbf{0.9137}$ \\
Nonuniform-intensity interlayer & WPGS
    & $\mathbf{0.9152}$ & $\mathbf{0.9231}$ & $\mathbf{0.9228}$ \\
\end{tabular}
\end{ruledtabular}
\end{table}

Table~\ref{tab:sm-gaussian-illumination} shows that WPGS maintains a high
transient-intensity floor for all tested Gaussian illumination profiles. In
the 2D task, WPGS gives \(q_{\min}=0.872\)--\(0.881\), whereas WGS develops
substantially deeper minima. In the 3D task, WPGS maintains
\(q_{\min}\geq0.896\), while WGS gives \(q_{\min}<0.05\) for every tested
value of \(R\). The nonuniform-target WPGS calculation similarly retains
\(q_{\min}\geq0.915\).  The weak variation of \(q_{\min}\) across the tested values of \(R\),
together with the consistently narrow trap-resolved phase distribution,
shows that the suppression of inter-frame phase mismatch and the transient-intensity improvement are maintained across the tested Gaussian illumination profiles. The refresh-stability improvement is therefore retained under substantial Gaussian nonuniformity of the incident field.

\section{Motional-energy and survival diagnostics}
\label{sec:sm-atom-dynamics}

During a finite SLM refresh, trapped atoms evolve in the time-dependent optical potential generated between consecutive holograms. We evaluate the resulting atomic motion by propagating thermal \(^{87}\mathrm{Rb}\) ensembles through the refresh-resolved potentials of the $2D$ and $3D$ reconfiguration sequences. The calculation uses the Gaussian optical-dipole and classical-trajectory framework established for optical-tweezer transport~\cite{grimm2000optical,Lee2016Three-dimensional,Schymik2020Enhanced,Pichard2024Rearrangement}. Motional heating during transport and whole-transport survival provide common dynamical diagnostics for comparing WGS and WPGS.

During each finite refresh, the pixel-resolved spatial field is reconstructed using the exact transient-field expression in Eq.~\eqref{eq:transient-field-exact}. The reconstructed field determines the instantaneous trap position \(\mathbf R_n(t)\), normalized depth \(q_n(t)\), and the occurrence of resolved trap splitting. When the local field remains single-peaked, the corresponding atomic potential is represented by a separable local approximation to the focused-Gaussian optical-dipole potential~\cite{grimm2000optical}:
\begin{equation}
V_n(\rho_n,z_n,t)
=
-U_0 q_n(t)
\frac{\exp[-2\rho_n^2/w_0^2]}
     {1+z_n^2/z_R^2},
\qquad
z_R=\frac{\pi w_0^2}{\lambda_{\rm tr}},
\label{eq:sm-gaussian-trap}
\end{equation}
where \(\rho_n\) and \(z_n\) are measured relative to \(\mathbf R_n(t)\). The field-derived \(q_n(t)\) and \(\mathbf R_n(t)\) retain the refresh-resolved evolution of the local depth and position, while \(w_0\) and \(z_R\) set the nominal radial and axial confinement scales.

We use \(^{87}\mathrm{Rb}\) atoms with
\(m=1.44316060\times10^{-25}~\mathrm{kg}\),
\(U_0/k_B=1.0~\mathrm{mK}\),
\(w_0=1.0~\mu\mathrm{m}\), and
\(\lambda_{\rm tr}=850~\mathrm{nm}\). These parameters give
\(z_R=3.70~\mu\mathrm{m}\),
\(\omega_r/2\pi=98.5~\mathrm{kHz}\), and
\(\omega_z/2\pi=18.8~\mathrm{kHz}\) in the harmonic limit~\cite{grimm2000optical}. Hologram propagation uses \(\lambda=820~\mathrm{nm}\), as specified in Sec.~\ref{sec:sm-task-definitions}.

Initial positions and velocities are sampled from the Maxwell--Boltzmann distribution at \(T_0=15~\mu\mathrm{K}\), with the position widths evaluated at the initial local trap depth~\cite{tuchendler2008energy}. We propagate \(N_{\rm MC}=512\) trajectories for every trap and use identical trap-resolved thermal samples for WGS and WPGS. Each \(\tau=10~\mu\mathrm{s}\) refresh is resolved by \(161\) time samples and integrated using the velocity-Verlet scheme~\cite{swope1982computer}. The motional excitation generated along each trajectory is quantified by the energy relative to the moving trap~\cite{guery2014transport}. For trajectory \(j\) associated with trap \(n\), we define
\begin{equation}
E_{{\rm mot},n,j}(t)
=
\frac{m}{2}
\left|
\dot{\mathbf r}_{n,j}(t)-\dot{\mathbf R}_n(t)
\right|^2
+
V_n\!\left(
\mathbf r_{n,j}(t)-\mathbf R_n(t),t
\right)
+
U_0q_n(t).
\label{eq:sm-motional-energy}
\end{equation}
The relative velocity gives the kinetic energy in the translating trap frame. Since \(V_n(0,0,t)=-U_0q_n(t)\), the additive term \(+U_0q_n(t)\) subtracts the instantaneous well-bottom energy, so that \(E_{{\rm mot},n,j}(t)\) measures the energy above the local trap minimum. Following energy-based analyses of excitation during transport in moving traps~\cite{guery2014transport}, we report the ensemble-averaged energy gain
\begin{equation}
\frac{\langle\Delta E_{\rm mot}\rangle}{k_BT_0}
=
\frac{
\left\langle
E_{\rm mot}(t_f)-E_{\rm mot}(0)
\right\rangle
}{
k_BT_0
}.
\label{eq:sm-energy-gain}
\end{equation}
Here \(k_BT_0\) is the characteristic energy scale of the initial Maxwell--Boltzmann ensemble used in the classical Monte Carlo sampling~\cite{tuchendler2008energy}. The same trajectories are used to evaluate whole-transport survival. A trajectory is classified as locally captured at time \(t\) when
\begin{equation}
\mathcal B_{n,j}(t)
=
\left\{
E_{{\rm mot},n,j}(t)-U_0q_n(t)<0,\;
\rho_{n,j}(t)<1.5w_0,\;
|z_{n,j}(t)|<1.5z_R
\right\},
\label{eq:sm-capture-condition}
\end{equation}
The whole-transport survival probability  for trap \(n\) is
\begin{equation}
P_n
=
\frac{1}{N_{\rm MC}}
\sum_{j=1}^{N_{\rm MC}}
\prod_k
\mathbf 1\!\left[\mathcal B_{n,j}(t_k)\right],
\label{eq:sm-survival}
\end{equation}
so that a trajectory contributes only when the capture condition is satisfied at every sampled time. Averaging over all traps gives
\begin{equation}
P
=
\frac{1}{N_{\rm trap}}
\sum_{n=1}^{N_{\rm trap}}P_n,
\qquad
N_{\rm loss}
=
\sum_{n=1}^{N_{\rm trap}}
\left(1-P_n\right)
=
N_{\rm trap}\left(1-P\right).
\label{eq:sm-survival-loss}
\end{equation}
For one initially occupied atom per trap, \(N_{\rm loss}\) gives the expected number of atoms that fail the whole-transport survival criterion in the finite Monte Carlo ensemble. Then, we apply the above dynamical model to the complete two- and three-dimensional reconfiguration sequences for both WGS and WPGS. The resulting atomic trajectories are evaluated as summarized in Table~\ref{tab:sm-pixel-dynamics}.

\begin{table*}[htbp]
\centering
\caption{\textbf{Pixel-resolved trap-splitting and atom-dynamics diagnostics.}
Trap splitting denotes the emergence of two resolved local intensity maxima
within a nominally single-tweezer region during refresh; a check mark indicates
that splitting occurs. Uncertainties are standard errors over traps. For the
three-dimensional task, the results are reported separately for the three
layers.}
\label{tab:sm-pixel-dynamics}
\begin{ruledtabular}
\begin{tabular}{llllccc}
Task &
Layer, \(z~(\mu\mathrm{m})\) &
Method &
Trap splitting &
\(\langle\Delta E_{\rm mot}\rangle/(k_BT_0)\) &
\(P\) &
\(N_{\rm loss}\) \\
\hline
\multirow{2}{*}{2D reconfiguration}
& \multirow{2}{*}{\(0\)}
& WGS
& \(\times\)
& \(0.5396\pm0.0093\)
& 0.999098
& 0.9238 \\

& & WPGS
& \(\times\)
& \(\mathbf{0.2789\pm0.0055}\)
& \(\mathbf{1.000000}\)
& \(\mathbf{0}\) \\
\hline
\multirow{6}{*}{3D reconfiguration}
& \multirow{2}{*}{\(-30\)}
& WGS
& \(\checkmark\)
& \(6.3054\pm0.1991\)
& 0.889038
& 113.6250 \\

& & WPGS
& \(\times\)
& \(\mathbf{0.4540\pm0.0067}\)
& \(\mathbf{1.000000}\)
& \(\mathbf{0}\) \\
\cline{2-7}
& \multirow{2}{*}{\(0\)}
& WGS
& \(\checkmark\)
& \(2.5618\pm0.0328\)
& 0.999098
& 0.9238 \\

& & WPGS
& \(\times\)
& \(\mathbf{0.1376\pm0.0019}\)
& \(\mathbf{1.000000}\)
& \(\mathbf{0}\) \\
\cline{2-7}
& \multirow{2}{*}{\(+30\)}
& WGS
& \(\checkmark\)
& \(3.3313\pm0.0373\)
& 0.992929
& 7.2402 \\

& & WPGS
& \(\times\)
& \(\mathbf{0.2048\pm0.0032}\)
& \(\mathbf{1.000000}\)
& \(\mathbf{0}\) \\
\end{tabular}
\end{ruledtabular}
\end{table*}

In the two-dimensional task, neither method produces optical-tweezer splitting, yet WPGS substantially reduces motional heating during transport, and no whole-transport survival failure is observed for WPGS in the finite Monte Carlo ensemble. The improvement therefore arises from stabilizing the trap depth and position throughout each refresh. The distinction becomes more pronounced in the three-dimensional task. Averaged over all \(3072\) traps, WPGS reduces the motional heating during transport from \(4.0662\,k_BT_0\) to \(0.2655\,k_BT_0\) and increases \(P\) from \(0.960355\) to \(1\). The layer-resolved results show that WGS produces optical-tweezer splitting in all three planes, with the strongest heating and \(93.3\%\) of its expected survival loss occurring in the \(z=-30~\mu\mathrm{m}\) layer. In contrast, no splitting is detected for WPGS, and all sampled WPGS trajectories satisfy the whole-transport survival criterion in every layer. Thus, the improved atomic confinement extends across the full three-dimensional transport geometry rather than being confined to a single layer.

\section{Computational cost of sequential hologram generation}
\label{sec:sm-runtime}

\begin{table}[htbp]
\centering

\caption{Computational cost of sequential hologram generation on a $1024\times1024$ phase-only SLM grid. The three tasks are the two-dimensional reconfiguration, simultaneous three-dimensional reconfiguration, and nonuniform-intensity interlayer transport described above. For each method, the reported iteration count is the minimum required to satisfy the same prescribed endpoint-convergence criterion and is also used for the corresponding results presented in the main text. ``Mean'' denotes the average wall-clock time per hologram update.}

\label{tab:sm-runtime}
\begin{ruledtabular}
\begin{tabular*}{0.82\textwidth}{@{\extracolsep{\fill}}lrrrr}
Task & WGS iter. & WGS mean (ms) & WPGS iter. & WPGS mean (ms) \\
\hline
2D reconfiguration & $26$ & $7.934$ & $5$ & $\mathbf{1.932}$ \\
3D reconfiguration & $26$ & $19.699$ & $5$ & $\mathbf{4.252}$ \\
Interlayer transport & $26$ & $5.348$ & $5$ & $\mathbf{1.258}$ \\
\end{tabular*}
\end{ruledtabular}
\end{table}

To quantify the computational overhead of generating successive holograms, we benchmark the wall-clock time per hologram update for the 2D, 3D, and nonuniform-intensity interlayer tasks. All benchmarks are performed on an NVIDIA A800 80GB PCIe GPU. We compare WPGS with the WGS-type procedure of Ref.~\cite{kim2019large}, using for each method the minimum iteration count required to satisfy the same prescribed endpoint-convergence criterion. We verified that these iteration counts are insensitive to whether random initialization or warm-start initialization is used. The same minimum iteration counts are used throughout the corresponding results presented in the main text. Table~\ref{tab:sm-runtime} reports the resulting mean wall-clock time per hologram update.
\end{document}